\documentclass[11pt]{article}

\usepackage{hyperref}
\pdfoutput=1

\usepackage{natbib}

\begin{document}
\title{Cicada: a Heavy but Agile Flyer}
\author{Kuo Gai, Yan Ren, Hui Wan, Haibo Dong\\
        \vspace{6pt} Department of Materials and Mechanical Engineering, \\
        Wright State University, Dayton, OH 45435, USA}
\maketitle
\begin{abstract}
``Cicada: a Heavy but Agile Flyer'' is a fluid dynamic video submitted to Gallery of Fluid Motion in APS-DFD 2011.
Comparing to other insects, cicadas can generate much higher lift to overcome their large body weight. The hidden mechanism may help in designing a Micro Air Vehicle (MAV) to carry large payloads. However, it is lack of literatures in discussing how cicadas use their wings to accomplish various flights. In this work, a high-speed photogrammetry system and 3D surface reconstruction technology are used to reveal cicada wing kinematics and deformation during a freely forward flight. The aerodynamic performance is studied using in-house immerse boundary method based Computational Fluid Dynamics(CFD) solver.
\end{abstract}
\bibliographystyle{jfm}
\bibliography{bib}

\end{document}